\documentclass[letterpaper,11pt]{article}

\usepackage{dynkin-diagrams}
\usepackage{cite}
\usepackage{amsmath}
\usepackage{amssymb}
\usepackage{mathrsfs}
\usepackage{xypic}
\usepackage{tikz-cd}

\usepackage{hyperref}
\usepackage{color}
\definecolor{darkred}{rgb}{0.65,0.15,0}
\hypersetup{pdfborder={0 0 0},colorlinks=true,urlcolor=darkred,citecolor=blue,linkcolor=darkred,linktocpage=true}

\usepackage{float} 

\usepackage{geometry}

\newgeometry{vmargin={35mm}, hmargin={35mm}} 

\usepackage{keyval}
\makeatletter
\define@key{setpar}{left}[0pt]{\leftmargin=#1}
\define@key{setpar}{right}[0pt]{\rightmargin=#1}
\define@key{setpar}{both}{\leftmargin=#1\relax\rightmargin=#1}
\makeatother

\makeatletter
\renewcommand*\env@matrix[1][\arraystretch]{%
  \edef\arraystretch{#1}%
  \hskip -\arraycolsep
  \let\@ifnextchar\new@ifnextchar
  \array{*\c@MaxMatrixCols c}}
\makeatother  
  
\def\eg{{\it e.g.}}
\def\ie{{\it i.e.}}

\def\adj{\hbox{\bf adj}}

\def\EWeight#1#2#3#4{\bigl({}^{\mathstrut}_{#1\mathstrut}{}_{#2\mathstrut}^{#4\mathstrut}{}_{#3\mathstrut}^{\mathstrut}\bigr)}

\def\DWeight#1#2#3{\bigl(\raise2.5pt\hbox{${}_{#1}$}{}^{#2}_{#3}\bigr)}

\def\AAWeight#1#2{\bigl(\raise0pt\hbox{${}^{#1}_{#2}$}\bigr)}

\def\fg{{\mathfrak g}}

\def\fe{{\mathfrak e}}
\def\so{{\mathfrak{so}}}
\def\sl{{\mathfrak{sl}}}

\def\nn{\nonumber}

\def\gl{\mathfrak{gl}}
\def\so{\mathfrak{so}}

\def\fk{\mathfrak{k}}

\def\*{\partial}

\def\transpose{\intercal}

\def\small#1{{\hbox{$#1$}}}
\def\fraction#1{\small{1\over#1}}
\def\fr{\fraction}

\def\CC{{\mathscr C}}

\def\RR{{\mathbb R}}

\def\LL{{\mathscr L}}

\def\Lie{\hbox{Lie}}

\begin{document}

\frenchspacing

\includegraphics[height=2cm]{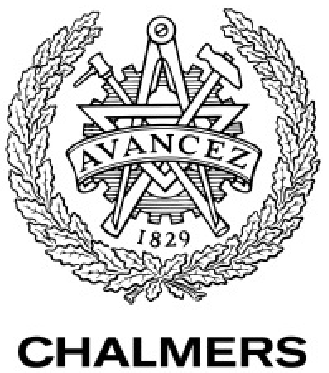}
\hspace{2mm}
\includegraphics[height=1.85cm]{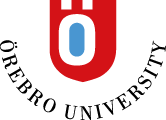}

\vspace{-12mm}
{\flushright Gothenburg preprint \\ 
}

\vspace{4mm}

\hrule

\vspace{16mm}


\thispagestyle{empty}

\begin{center}
  {\Large \bf \sc The teleparallel complex}
    \\[10mm]
    
{\large
Martin Cederwall${}^1$ and Jakob Palmkvist${}^{2}$}

\vspace{10mm}
       {\footnotesize ${}^1${\it Department of Physics,
         Chalmers Univ. of Technology,\\
 SE-412 96 Gothenburg, Sweden}}

\vspace{2mm}
       {\footnotesize ${}^2${\it Department of Mathematics,
         \"Orebro Univ.,\\
 SE-701 82 \"Orebro, Sweden}}

\end{center}

\vfill

\begin{quote}
\textbf{Abstract:} 
We formalise the teleparallel version of extended geometry (including gravity) by the introduction of a complex, the differential of which provides the linearised dynamics. The main point is the natural replacement of the two-derivative equations of motion by a differential which only contains terms of order 0 and 1 in derivatives. Second derivatives arise from homotopy transfer (elimination of fields with algebraic equations of motion). 
The formalism has the advantage of providing a clear consistency relation for the algebraic part of the differential, the ``dualisation'', which then defines the dynamics of physical fields. It remains unmodified in the interacting BV theory, and the full non-linear models arise from covariantisation.
A consequence of the use of the complex is that symmetry under local rotations becomes as good as manifest, instead of arising for a specific combination of tensorial terms, for less obvious reasons.
We illustrate with a derivation of teleparallel Ehlers geometry, where the extended coordinate module is the adjoint module of a finite-dimensional simple Lie group. 
\end{quote} 

\vfill

\hrule

\noindent{\tiny email:
  martin.cederwall@chalmers.se, jakob.palmkvist@oru.se}

\newpage

\tableofcontents

\section{Introduction}

Extended geometry
\cite{Cederwall:2017fjm,Cederwall:2018aab,Cederwall:2019qnw,Cederwall:2019bai,Cederwall:2021xqi,Bossard:2023ajq}
is a general framework for the construction of gravitational theories, that replaces the structure group $GL(d)$ of gravity with an arbitrary semisimple group $G$ and $d$-dimensional vectors with elements in some module of $G$.
Coordinates for an extended space belong to this module, but a ``section constraint'' restricts coordinate dependence locally to a $GL(d)$ vector.
Special cases of extended geometry are double geometry 
\cite{Tseytlin:1990va,Siegel:1993xq,Siegel:1993bj,Hitchin:2010qz,Hull:2004in,Hull:2006va,Hull:2009mi,Hohm:2010jy,Hohm:2010pp,Jeon:2012hp,Park:2013mpa,Berman:2014jba,Cederwall:2014kxa,Cederwall:2014opa,Cederwall:2016ukd} 
and exceptional geometry
\cite{Hull:2007zu,Pacheco:2008ps,Hillmann:2009pp,Berman:2010is,Berman:2011pe,Coimbra:2011ky,Coimbra:2012af,Berman:2012vc,Park:2013gaj,Cederwall:2013naa,Cederwall:2013oaa,Aldazabal:2013mya,Hohm:2013pua,Blair:2013gqa,Hohm:2013vpa,Hohm:2013uia,Hohm:2014fxa,Cederwall:2015ica,Butter:2018bkl,Bossard:2017aae,Bossard:2018utw,Bossard:2019ksx,Bossard:2021jix,Bossard:2021ebg}, and of course gravity itself.

It has become abundantly clear 
\cite{Palmkvist:2015dea,Cederwall:2017fjm,Cederwall:2018aab,Cederwall:2019qnw,Cederwall:2019bai,Cederwall:2021xqi}
that tensor hierarchy algebras (THA's)
\cite{Palmkvist:2013vya,Carbone:2018xqq,Cederwall:2019qnw,Cederwall:2021ymp,Cederwall:2022oyb}
provide a foundation for extended geometry.
They do so in providing a clearly defined set of fields, ghosts etc., which are given by the content of the relevant THA
at definite degrees in certain gradings. 
One way of viewing this structure is that the THA's provide the structure for extended geometry which is the analogue of forms and extxerior derivatives for Yang--Mills (YM) theory (or higher form gauge theory).
This is the point of view which is taken in the present paper. 
In order to define YM on some manifold, it is not enough to know the complex of forms, with its exterior derivative. One also needs to introduce the two-derivative dynamics defined by dualisation in terms of the ``kinetic operator'' $d\star d$.
The situation in gravity and extended geometry turns out to be analogous. We will present a complex for extended geometry containing a 
``dualisation'', which indeed gives full information concerning the dynamics.

The formalism presented completes the construction of extended geometry from tensor hierarchy algebras, in the sense that the correspondence between elements in the tensor hierarchy algebra and a full set of Batalin--Vilkovisky (BV) fields for the extended geometry is given.
The question of finding the dynamics is a question of constructing the complex, in particular of finding the ``dualisation'', which is part of the differential on the complex. This is an algebraic problem.
We hope that this clarification will be useful when dealing with infinite-dimensional structure groups.

We begin by discussing in Section \ref{sec:teleparallel}, in some detail, the dualisations used to "turn the complex around" in YM theory and in (teleparallel) gravity. In the latter case we put focus on the consistency relation for the dualisation.
This paves the way for the construction of the teleparallel complex for extended geometry in Section
\ref{sec:extended}, which also contains some background on extended geometry needed in the construction.
Section \ref{sec:examples} specialised to classes of examples, in particular ``Ehlers extended geometry'', where the coordinate module is the adjoint of a semisimple finite-dimensional Lie group.
Section \ref{sec:BV} provides the structure of the ``central'' part of the BV action, involving fields and antifields.
We conclude with a discussion of open issues, in particular the application of the present results to 
infinite-dimensional structure groups.

\section{The teleparallel complex\label{sec:teleparallel}}

\subsection{Turning a complex around---the Yang--Mills example\label{sec:YMcomplex}}

Sometimes fields (including ghosts) naturally belong to a complex with a differential linear in derivatives, such as the de Rham differential, but the equations of motion for the physical fields are second order in derivatives. This can be formalised by a ``turning around'' of the complex, or a ``doubling'', where antifields belong to the dual complex.

A main example is Yang--Mills theory on a manifold $M$, where ghosts are in $\Omega^0(M)$ and connections in $\Omega^1(M)$. Antifields are in the dual spaces. The turning around of the complex is acheived by 
connecting the original complex to the dual one by a ``duality'' operation $\sigma$, not containing derivatives
\cite{Zeitlin:2008cc,Rocek:2017xsj,Reiterer:2019dys}.
\begin{equation}
\label{YMComplex}
    \begin{tikzcd}[row sep = 16 pt, column sep = 16 pt]
    \hbox{ghost\#}=&1&0&-1&-2\\
       & \Omega^0\ar[r,"d"]&\Omega^1\ar[r,"d"]&\Omega^2 \\
        &&\Omega^{d-2}\ar[r,"d",swap]\ar[ur,"\sigma" near start]&\Omega^{d-1}\ar[r,"d",swap]
        &\Omega^d
\end{tikzcd}
\end{equation}
Here, the differential is $q=d+\sigma$, and $\sigma=\star$.
Note that the forms in the upper line are truncated at the point where the complex turns. 
There are no ghost antifields in $\Omega^3$, nor any ghosts in $\Omega^{d-3}$.   
Let us call this complex $\CC_{\hbox{\tiny YM}}$.
We will use it to introduce some concepts and procedures which will be of use later, when we turn to gravity and extended geometry. 

The complex comes equipped with a natural pairing (wedge product and integration) between elements on the upper and lower lines.
It agrees with the Batalin--Vilkovisky (BV) pairing, carrying ghost number $1$, of fields with their antifields. 
Elements in the complex are sets of BV fields with the appropriate parity: bosonic for even ghost number, fermionic for odd.
A scalar product $\langle\cdot,\cdot\rangle$ on elements $\Psi=\psi+\bar\psi$, where $\psi$ belongs to the upper line and $\bar\psi$ to the lower, is defined so that 
\begin{gather}
\langle\psi,\psi'\rangle=0=\langle\bar\psi,\bar\psi'\rangle\;,\nn\\
\langle\psi,\bar\psi\rangle=\int_M\psi\wedge\bar\psi=\langle\bar\psi,\psi\rangle\;.
\end{gather}
The linearised BV action is 
\begin{align}
S=\fr2\langle\Psi,q\Psi\rangle\;.
\end{align}

The cohomology is linearised Yang--Mills. 
Let us check the action for the physical fields. Denote the physical 1-form in the upper line $A$ and the 2-form antifield $F$,
and their antifields $\bar A$ and $\bar F$. Note that $F$ is not the field strength of $A$, but an independent antifield, and that $\bar F$ is a physical field (ghost number 0).
We can now evaluate the relevant part of the action (taking $\*M=\emptyset$ for simplicity) as
\begin{align}
S_0[A,\bar F]&=\fr2\left(
\langle A,d\bar F\rangle+\langle\bar F,dA+\sigma\bar F\rangle\right)\nn\\
&=\langle\bar F,F(A)\rangle+\fr2\langle\bar F,\sigma\bar F\rangle\;,
\end{align}
where $F(A)=dA$.
Solving the algebraic equations of motion obtained by varying $\bar F$ gives 
$\bar F=-\sigma^{-1} F(A)$, and reinserting in the action yields the standard two-derivative action
\begin{align}
S'_0[A]=-\fr2\langle F(A),\sigma^{-1}F(A)\rangle\;.
\end{align}

This elimination of $\bar F$ can equivalently be expressed as
homotopy transfer to the cohomology of $\sigma$.
This is done by first constructing a strong homotopy retract
to the cohomology of $\sigma$, which we call $\CC'_{\hbox{\tiny YM}}$, with zero differential,
\begin{equation}
    \begin{tikzcd}
(\CC'_{\hbox{\tiny YM}},0) \arrow[r, shift left=1ex, "{i}"] 
 &  \arrow[l, shift left=1ex, "{p}"]  (\CC_{\hbox{\tiny YM}},\sigma) \arrow[loop right, distance=3em, start anchor={[yshift=1ex]east}, end anchor={[yshift=-1ex]east}]{}{h}
\end{tikzcd}\;,
\end{equation}
where the inclusion $i$ and the projection $p$ are the na\"\i ve ones, identifying the spaces in $\CC'_{\hbox{\tiny YM}}$ with the ones in $\CC_{\hbox{\tiny YM}}$, and $h=\sigma^{-1}$.
Since $\sigma$ is invertible, the new complex $\CC'_{\hbox{\tiny YM}}$ consists of the vector spaces in $\CC_{\hbox{\tiny YM}}$, with $F\in\Omega^2$ and
$\bar F\in\Omega^{d-2}$ removed.
The horizontal differential $d$ is the seen as a perturbation of $\sigma$, and the homological perturbation lemma
\cite{Lapin} yields the quasi-isomorphism
\begin{equation}
    \begin{tikzcd}
(\CC'_{\hbox{\tiny YM}},q') \arrow[r, shift left=1ex, "{i'}"] 
 &  \arrow[l, shift left=1ex, "{p'}"]  (\CC_{\hbox{\tiny YM}},q=\sigma+d) \arrow[loop right, distance=3em, start anchor={[yshift=1ex]east}, end anchor={[yshift=-1ex]east}]{}{h'}
\end{tikzcd}\;,
\end{equation}
where the new differential $q'$ (and also $h'$, $i'$ and $p'$) is given as a perturbation series in $d$,
\begin{align}
q'=\sum_{n=0}^\infty p(dh)^ndi\;.
\end{align}
Here, only the terms with $n=0,1$ contribute, and dropping $p$ and $i$ (since they are given by the trivial identification of vector spaces), we obtain
\begin{align}
q'=d+d\sigma^{-1}d\;,
\end{align}
just like the result of algebraic elimination of fields above.
\begin{equation}
    \begin{tikzcd}[row sep = 16 pt, column sep = 16 pt]
        \Omega^0\ar[r,"d"]&\Omega^1\ar[dr,out=0,in=180,looseness=4,"d\sigma^{-1}d", near start]\\
        &&\bar\Omega^{d-1}\ar[r,"d",swap]
        &\bar\Omega^d
\end{tikzcd}
\end{equation}

The interacting YM theory is obtained by
``covariantisation'', which formally means Chern--Simons theory on the complex. The operator $\sigma$ of course remains undeformed.

Note that the physical input in the models lies entirely in the choice of $\sigma$. It is of course completely specified by a choice of metric on $M$. There is no other consistency imposed on $\sigma$ than its invertibility.
One might think that the use of the complex \eqref{YMComplex} just introduces an unnecessary complication in a known 
model. However, the interactions become of lower order (essentially Chern--Simons), and the dynamical input is concentrated in a linear operator, which stays unmodified in the non-linear Batalin--Vilkovisky theory.
The last property will be valuable in gravity and extended geometry, where the ``duality'' operation is subject to 
a consistency relation.
We also notice that this mechanism is at work in supersymmetric pure spinor field theory
\cite{Cederwall:2011vy,Cederwall:2009ez,Cederwall:2010tn,Cederwall:2013vba,Cederwall:2022fwu,Eager:2021wpi}, where it is responsible for the simple forms of interactions, which are generically of lower polynomial order than the interaction terms for component fields.

\subsection{The teleparallel gravity complex}

We will now construct the analogous turning around of the linear complex for the teleparallel version of gravity
(see \eg\ refs. \cite{DeAndrade:2000sf,Cederwall:2021xqi,Golovnev:2023yla}).
The main structure will carry over directly to extended geometry.
The interesting behaviour that distinguishes these models from the 
Yang--Mills complexes is that there is a consistency relation on the ``dualisation'', the operator corresponding to
$\sigma$ in the complex \eqref{YMComplex}. As we will see, it arises simply from $q^2=0$, $q$ being the differential on the complex.
An advantage is that this relation can be solved for $\sigma$ already in the linear theory, \ie, all the defining structure lies in the construction of a $1$-bracket. Higher brackets will arise as covariant ``decorations'' of the $1$-bracket.

In teleparallel gravity in $d$ dimensions, the infinitesimal diffeomorphism symmetry is parametrised by vector fields (we denote the space of vector fields $V$), but there is also a local Lorentz symmetry.
The physical fields normally consist of a vielbein, which linearly around some background is represented locally as an element $e$ in
$\gl(d)$, or in $\Omega^1\otimes V$. The field strength is the torsion $\theta(e)$ of the Weitzenb\"ock connection, which
belongs to $\Omega^2\otimes V$. 
Inspired by the YM complex \eqref{YMComplex}, we put the diffeomorphism parameters $v$, the vielbein $e$ and the ``torsion'' $\theta$ in the upper line, and their duals in the lower line (note that $\theta$ is not $\theta(e)$ but an independent antifield in the torsion module).
Dual fields here means dual with respect to integration, so the lower line will consist of tensor densities.

Where does the local Lorentz ghosts fit in? 
It is well known that the torsion $\theta(e)$, while being a tensor under diffeomorphisms, is not invariant under local Lorentz transformations. Instead the transformation of the torsion is the covariant derivative of the Lorentz parameter \cite{Cederwall:2021xqi}.  This implies that a differential 
$\bar h\rightarrow e\rightarrow\theta$, $\bar h$ representing the Lorentz ghosts, does not square to $0$. 
Still these arrows must be part of the complex, and the only solution is that it there is another path from $\bar h$ to $\theta$, now through $\bar\theta$, the other field of ghost number $0$, We then have arrows
\begin{equation}\label{Gsubdiagram}
    \begin{tikzcd}[row sep = 16 pt, column sep = 16 pt]
        &e\ar[r]&\theta \\
        \bar h\ar[ur]\ar[r]
        &\bar\theta\ar[ur]
        \end{tikzcd}
\end{equation}
The part $\sigma$ of the differential from $\bar\theta$ to $\theta$ is analogous to the dualisation in the first Yang--Mills complex, but it will now be subject to a relation
\begin{align}
\boxed{
\;d\circ\varrho+\sigma\circ d=0\;,\;\label{eq:keypoint}
}
\end{align}
where $\varrho$ represents the action of the linearised Lorentz transformations (the left $\nearrow$ in \eqref{Gsubdiagram}).
This is the key point, and the principle carries over to extended geometry.
Note that arrows $\rightarrow$ are linear in derivatives, while arrows $\nearrow$ are algebraic.

Since the ghosts $\bar h$ are now introduced, also their antifields of ghost number $-2$ must be present, and we see that they will sit in the upper line in the position of torsion Bianchi identies. The latter are in 
$\Omega^3\otimes V$, containing a contraction $\Omega^2$, which can represent an element in the local Lorentz subalgebra.
Unlike the Yang--Mills case, where Bianchi identities were left out in the upper line, the correct turning around of the gravity complex demands that we keep a submodule of the torsion Bianchi identites as ghost antifields.
We end up with a tentative complex:
\begin{equation}\label{Gdiagram}
    \begin{tikzcd}[row sep = 16 pt, column sep = 16 pt]
    \hbox{ghost\#}=&1&0&-1&-2\\
        &V\ar[r,"d"]&\Omega^1\otimes V\ar[r,"d"]&\Omega^2\otimes V\ar[r,"d"]&\Omega^2 \\
        &\overline{\Omega^2}\ar[r,"d",swap]\ar[ur,"\varrho" near start]
        &\overline{\Omega^2\otimes V}\ar[r,"d",swap]\ar[ur,"\sigma" near start] 
        &\overline{\Omega^1\otimes V}\ar[r,"d",swap]\ar[ur,"\varrho^\ast" near start]&\overline V
\end{tikzcd}
\end{equation}
The vector spaces in the lower line are dual under integration to the ones in the upper line; they are densities with covariant naked divergences. 
The action of the horizontal derivatives is
\begin{align}
d\left(\begin{matrix}
	v^m&e_m{}^n&\theta_{mn}{}^p&h_{mn}\\
	\bar h^{mn}&\bar\theta_m{}^{np}&\bar e_m{}^n&\bar v_m
\end{matrix}\right)
=\left(\begin{matrix}
	0&\*_mv^n&2\*_{[m}e_{n]}{}^p&3\*_{[m}\theta_{np]}{}^p\\
	0&3\delta_m^{[n}\*_q\bar h_{\mathstrut}^{pq]}&2\*_p\bar\theta_m{}^{np}&\*_n\bar e_m{}^n
\end{matrix}\right)
\end{align}
The $\varrho$ arrows are also obvious, they are given as
$(\varrho\bar h)_m{}^n=g_{mp}\bar h^{pn}$ and $(\varrho^*\bar e)_{mn}=\bar e_{[m}{}^pg_{n]p}$.
$g$ is some ``background metric'' invariant under an $\so(d)$ subalgebra of some signature, which for convenience is taken to carry the correct weight to transform from densities to tensors
(in the non-linear theory it is of course obtained from the vielbein).
The horizontal derivatives in the lower row are given by the same structure constants as the ones in the upper row.
It is straightforward to verify that the identity from the right parallellogram,
$d\sigma+\varrho^*d=0$, is equivalent to eq. \eqref{eq:keypoint} as long as $\sigma$ is symmetric (as a matrix in indices for the torsion module). This is a direct consequence of the diagonal action of $\sigma$ on $\so(d)$-modules.

Now, only $\sigma$ needs to be constructed.
It is convenient to lower all indices on $\theta$ and $\bar\theta$ with the metric and to parametrise $\sigma$ in terms of endomorphisms on $\Omega^1\otimes\Omega^2$. Let $X\in\Omega^1\otimes\Omega^2$  be represented by a tensor $X_{m,np}$ antisymmetric in the last pair. The endomorphisms are then spanned by $\{1,\alpha,\beta\}$, where
\begin{align}
(\alpha X)_{m,np}&=X_{[n,p]m}\;,\nn\\
(\beta X)_{m,np}&=g_{m[n}g^{qr}X_{|q|,p]r}\;.
\end{align}
They fulfill 
\begin{align}
\alpha^2&={1\over2}(1+\alpha)\;,\nn\\
\alpha\beta=\beta\alpha&=-{1\over2}\beta\;,\\
\beta^2&=-{d-1\over2}\beta\;.\nn
\end{align}
Let $X_{m,np}=g_{nn'}g_{pp'}\*_m\bar h^{n'p'}$. 
Then, one has one side of the left parallellogram as
$d\varrho\bar h=2\alpha X$. We also have $d\bar h=(1+2\beta)X$. To satisfy $(d\varrho+\sigma d)\bar h=0$ we are uniquely led to
\begin{align}
\sigma=-2\alpha(1+2\beta)^{-1}=-2\alpha(1+{2\over d-2}\beta)=-2\alpha+{2\over d-2}\beta\;.
\end{align}
The inverse map is 
\begin{align}
\sigma^{-1}={1\over2}-\alpha+2\beta\;.\label{eq:sigmainverse}
\end{align}

The classical action, \ie, the part of the linearised BV action containing only physical fields of ghost number $0$, is read off from the complex as
\begin{align}
S_0[e,\bar\theta]=\fr2\left(
\langle \bar\theta,de\rangle + \langle e,d\bar\theta\rangle + \langle \bar\theta,\sigma\bar\theta\rangle
\right)
=\langle e,d\bar\theta\rangle+\fr2\langle \bar\theta,\sigma\bar\theta\rangle\;.
\end{align}
The equations of motion for $\bar\theta$ are algebraic, and solved by $\bar\theta=-\sigma^{-1}de$.
Note the close analogy to the Yang--Mills case.
Reinserting this solution in $S_0$ gives the linearised action of standard teleparallel gravity
\begin{align}
S'_0[e]=-\fr2\langle de,\sigma^{-1}de\rangle=-\fr2\langle\theta(e),\sigma^{-1}\theta(e)\rangle,
\end{align}
where $\theta(e)=de$ is the linearised torsion and the ``dualisation'' $\sigma^{-1}$ is given by
eq. \eqref{eq:sigmainverse}.
This procedure can of course also be described as homotopy transfer as in Section \ref{sec:YMcomplex}.
The full non-linear action is only a matter of covariantisation; this will be described in
Section \ref{sec:BV}.

\section{Extended geometry complexes\label{sec:extended}}

Now, we want to apply the same construction to extended geometry. The principle is the same as for gravity. The Bianchi identity in the antisymmetric module plays a special r\^ole, leading to the ``dual gauge symmetry'' of ref. \cite{Cederwall:2021xqi}.
There, it was realised that the Bianchi identity, projected to the local subalgebra, was responsible for the local symmetry. The present paper provides the explanation.

\subsection{Fields from tensor hierarchy algebras}

The version of extended geometry we need to rely on is clearly the teleparallel one 
\cite{Cederwall:2021xqi}. It is adapted to the identification of gauge parameters, fields, torsion (field strengths) and Bianchi identities from the THA. 
The structure algebra is $\fg=\Lie(G)$. In the present paper, $\dim\fg<\infty$.
The coordinate module is a lowest weight $\fg$-module $R(-\lambda)$. The tensor hierarchy algebra used for the construction is
$S(\fg^+)$, a non-contragredient super-extension of the Lie algebra $\fg^+$, in turn obtained by ``extending $\fg$ at $\lambda$''. If $\lambda$ is a fundamental weight this simply means adding a node to the Dynkin diagram of $\fg$, singly connected to the node of $\lambda$. 
We refer to refs. \cite{Cederwall:2019qnw,Cederwall:2019bai,Cederwall:2021ymp} for notation and full detail.
Since the use of the tensor hierarchy algebra for identification of all fields (and for deriving the form of the differential) is an integral part of our program, we review it briefly. This part can be skipped, if one accepts the assignments.

\begin{figure}
  \centerline{
  \includegraphics{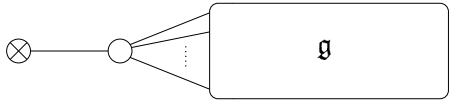}}
  \caption{\it The Dynkin diagram for 
  the tensor hierarchy algebras
  $W(\fg^+)$
    and $S(\fg^+)$. Removing the ``grey'' node in the right diagram yields the Dynkin diagram of $\fg^+$.}
  \label{fig:Dynkin}
\end{figure}

The tensor hierarchy algebra $S(\fg^+)$ can be constructed from a Dynkin diagram where a grey node is added
to the diagram of $\fg^+$ (Figure~\ref{fig:Dynkin}), and then comes with a bigrading, where a $\fg$-module appears for any pair of integers,
which can be chosen in different ways.
We here call them $p$ and $q$ and define the bigrading such that the $e$-generators associated to the grey node and the node next to it sit at
$(p,q)=(0,-1)$ and $(p,q)=(1,1)$, respectively. Then tensor hierarchy algebra $W(\fg)$ is a subalgebra at $q=0$.
By a further decomposition of it into $\fg$-modules at different $p$,
non-ancillary fields of ghost number $\gamma$ are found at $(p,q)=(\gamma,0)$. In particular, we have generalised diffeomorphism parameters
in the coordinate module $R(-\lambda)$ at $(p,q)=(1,0)$, (linearised) physical fields---vielbeins---in $\fg\oplus\RR$ at $(p,q)=(0,0)$,
torsion at $(p,q)=(-1,0)$ and torsion Bianchi identities at $(p,q)=(-2,0)$.
In the spirit of Section \ref{sec:teleparallel}, the torsion module is populated by ``torsion antifields'' and part of the Bianchi identity module by ghost antifields for local rotations.

In addition there may be ancillary fields of various ghost numbers, found at $(p,q)=(\gamma-1,1)$.
For finite-dimensional $\fg$, ancillary fields are restricted to $\gamma\geq1$.
Ancillary fields arise due to the presence of tensors of mixed symmetry in a decomposition into modules of
$\gl(d)\subset\fg$, and can be seen as compensating for a failure of the Poincar\' e lemma for the derivative. They do not play a decisive r\^ole in the construction in the present paper---in particular they are irrelevant for the construction of the dualisation $\sigma$---and will largely be ignored.

In the following, we will need the fields and modules:
\begin{itemize}
\item generalised diffeomorphism parameters (ghosts) in $R(-\lambda)$, which we denote $V^M$;
\item the generalised vielbein $E_M{}^A$, which linearised around some background is para\-metrised by an element $E^\alpha$ in $\fg$ and a scalar $e$;
\item torsion (antifields) $\Theta_M{}^\alpha$ and $\theta_M$  in a specific set of modules, see below;
\item torsion Bianchi identities (ghost antifields) in an antisymmetric module $\wedge^2 R(\lambda)$.
\end{itemize}
There is always a ``small'' torsion $\theta_M$ in $R(\lambda)$.
The modules for the ``big'' torsion $\Theta_M{}^\alpha$ were characterised in ref. \cite{Cederwall:2019qnw}.
They are the modules $\Theta\subset R(\lambda)\otimes\fg$ that automatically respect the ideal $R(-2\lambda)$ of $W(\fg)$ at level $2$, so that $\Theta\otimes R(-2\lambda)\not\supset R(-\lambda)$.
 These turn out to be (we restrict to $\lambda$ being a fundamental weight, for simplicity)
 highest weight modules $R(\lambda+\gamma)$, where $\gamma$ is a highest 
 root with $(\lambda,\gamma)=
 0,-2,-3,\ldots,-(\lambda,\vartheta)$, where $\vartheta$ is the highest root of $\fg$ and the inner product is normalised by $(\vartheta,\vartheta)=2$.
 It is convenient to use an invariant tensor $\varphi^\alpha{}_{M,\beta}{}^N$, introduced in ref. \cite{Cederwall:2019qnw}, which is a weighted sum of projection operators on the $\Theta$-modules, and which occurs naturally as structure constants in the THA.
 It fulfills various identities, of which the most important is
 \begin{align}
 t_{\beta\langle M}{}^P\varphi^\beta{}_{N\rangle,\alpha}{}^Q=0\;,
 \label{eq:phiidentity}
 \end{align}
$\langle MN\rangle$ denoting projection on $R(2\lambda)$, manifesting that the level $2$ ideal is respected.

\subsection{The differential\label{sec:differential}}

The identification of fields above gives the extended geometric analogue of forms for YM theory, or of the upper line of the gravity complex \eqref{Gdiagram}. There is a differential $d$, which between non-ancillary fields contains one derivative, in a contraction with
the adjoint action of elements in $S(\fg^+)$.
(There is also a part of $d$, mapping ancillary to non-ancillary fields, acting by embedding, which contains no derivatives, and a $1$-derivative part mapping ancillary to ancillary.) 
Besides algebraic relations in $S(\fg^+)$,
the nilpotency $d^2=0$ relies on the section constraint
\begin{align}
&Y(\*\otimes\*)=0\;,\label{eq:sectionconstraint}
\\
&\varsigma Y=-\eta^{\alpha\beta}t_\alpha\otimes t_\beta+(\lambda,\lambda)-1+\varsigma\;,\nn
\end{align}
where $\eta$ is the inverse Killing metric and $\varsigma$ the permutation operator.

\subsubsection{Generalised diffeomorphisms}

Generalised diffeomorphisms are given by the generalised Lie derivative. The action on a covector of weight $w$ is
\begin{align}
\LL_V W_M=V^N\*_NW_M+\eta^{\alpha\beta}(t_\alpha\otimes t_\beta)_{MN}{}^{QP}\*_PV^NW_Q
+w\*_NV^NW_M\;.
\label{eq:gendiff}
\end{align}
The value of $w$ for a tensor is $w=1-(\lambda,\lambda)$. 
The closure of the commutator of generalised diffeomorphisms to generalised diffeomorphisms (and, when they are present, ancillary transformations) relies on the section constraint \eqref{eq:sectionconstraint}.

\subsubsection{Generalised vielbein and involution}

The generalised vielbein is a group element $E_M{}^A$ in $G\times\RR$, transforming from the right with the locally realised subgroup $K(G)$. The local subgroup $K(G)$, with Lie algebra $\fk$, is thought of as the maximal compact subgroup, more generally an involutory subgroup.

The choice of $\fk\subset\fg$ is parametrised by a generalised metric $G_{MN}$ on $R(-\lambda)$. 
In the non-linear theory the involution is of course identified dynamically as $G_{MN}=(EHE^\transpose)_{MN}$, 
$H_{AB}$ being a positive definite $\fk$-invariant metric on $R(-\lambda)$. In the linearised theory around $E=1$, we can identify $G=H$.
The generalised metric defines an involution $\tau$ through
its action on the representation matrices $t_{\alpha M}{}^N$:
\begin{align}
\tau(t_\alpha)=-Gt_\alpha^\transpose G^{-1}\;.
\end{align}
The eigenvalue of $\tau$ is $+1$ on the local subalgebra $\fk$ and $-1$ on its complement $\fk^\perp$ with respect to the Killing metric. 
Accordingly, splitting $\alpha=(a,a')$,
$\tau(T_a)=T_a$, $\tau(T_{a'})=-T_{a'}$,
where $T_a\in\fk$, $T_{a'}\in\fk^\perp$.
Freely raising and lowering adjoint indices with the Killing metric $\eta_{\alpha\beta}$ and its inverse, the involution may be identified with (minus) the metric in the adjoint representation. $\tau(T_\alpha)=\tau_\alpha{}^\beta T_\beta$ gives
\begin{align}
\tau(T^\alpha)=-\tau^{\alpha\beta}T_\beta=G^{\alpha\beta}T_\beta\;.
\end{align}
$G_{\alpha\beta}=-\langle T_\alpha,\tau(T_\beta)\rangle$ is then positive definite. In the splitting $\alpha=(a,a')$, it is by definition
$G_{ab}=-\eta_{ab}$, $G_{a'b'}=\eta_{a'b'}$.
One can covariantly extract $\fk$ from an antisymmetric matrix $A_{MN}$ as
\begin{align}
H_a=(G^{-1}t_a)^{MN}A_{MN}\;.
\end{align}

Using the transformation \eqref{eq:gendiff} on the vielbein,
\begin{align}
\LL_V E_M{}^A=V^N\*_NE_M{}^A+\eta^{\alpha\beta}(t_\alpha\otimes t_\beta)_{MN}{}^{QP}\*_PV^NE_Q{}^A
+w\*_NV^NE_M{}^A\;,
\label{eq:Etransformation}
\end{align}
and linearising around $E=1$ as $E=1+t_\alpha E^\alpha+we$, we obtain
\begin{align}
\LL_VE^\alpha&=t^\alpha{}_M{}^N\*_NV^M\;,\nn\\
\LL_Ve&=\*_MV^M\;.
\end{align}
This is identified as the $1$-bracket $dV$.

The transformation of $E$ under a local transformation $H\in\fk$ is
$\delta_HE=EH$, which linearised becomes the (algebraic) $1$-bracket
as $\delta_HE^a=H^a$, $\delta_HE^{a'}=0$, $\delta_He=0$.


\subsubsection{Torsion}

The (generalised ) torsion of a vielbein $E$ is defined as the covariantly transforming part (as $\fg$-modules) of the 
Weitzenb\"ock connection (Maurer--Cartan form) $\Gamma_M=-\*_MEE^{-1}$.
Note that the Weitzenb\"ock connection, unlike other metric-compatible connections contains a derivative in its first index, and thus satisfies the section constraint (together with itself or with derivatives).
Writing $\Gamma_{MN}{}^P=t_{\alpha N}{}^P\Gamma_M{}^\alpha+w\delta_N^P\gamma_M$
and using the transformation of the vielbein, we obtain
the inhomogeneous parts of the transformation of the connection under generalised diffeomorphisms
(denoted by $\Delta_V$, the rest is the generalised Lie derivative),
\begin{align}
\Delta_V\Gamma_M{}^\alpha&=-t^\alpha{}_N{}^P\*_M\*_PV^N\;,\nn\\
\Delta_V\gamma_M&=-\*_M\*_NV^N\;.
\end{align}
We find the ``big'' and ``small'' torsion as
\begin{align}
\Theta_M{}^\alpha&=\varphi^\alpha{}_{M,\beta}{}^N\Gamma_N{}^\beta\;,\nn\\
\theta_M&=t_{\alpha M}{}^N\Gamma_N{}^\alpha-(\lambda,\lambda)\gamma_M\;.
\label{eq:torsionexpressions}
\end{align}
where the the identity \eqref{eq:phiidentity} for the $\varphi$ tensor is responsible for making the big torsion
$\Theta_M{}^\alpha$ covariant. The small torsion $\theta_M$ becomes covariant thanks to the symmetric section constraint.

Linearisation of eq. \eqref{eq:torsionexpressions} gives the $1$-bracket from $E=(E^\alpha,e)$ to the torsion modules as
\begin{align}
(dE)_M{}^\alpha&=-\varphi^\alpha{}_{M,\beta}{}^N\*_NE^\beta\;,\nn\\
(dE)_M&=-t_{\alpha M}{}^N\*_NE^\alpha+(\lambda,\lambda)\*_Me\;.
\label{eq:dEtoTheta}
\end{align}

\subsubsection{Bianchi identities}

In the cases considered in ref. \cite{Cederwall:2021xqi}, it was shown that
there is a Bianchi identity in the leading antisymmetric module, \ie, the highest module in $\wedge^2R(\lambda)$,
and it was also shown generally that this module is always present at $(p,q)=(-2,0)$.
This Bianchi identity can be observed in the tensor hierarchy algebra $S(\fg^+)$ in all examples we have checked, 
and we think it can be proven in general, using the approach proposed in ref. \cite{Cederwall:2022oyb}.
It takes the form
\begin{align}
t_{\alpha\{M|}{}^P(D_P+\theta_P)\Theta_{|N\}}{}^\alpha+2(2-{1\over(\lambda,\lambda)})D_{\{M}\theta_{N\}}=0\;.
\end{align}
Here, $D$ is the covariant derivative with the Weitzenb\"ock connection,
and $\{MN\}$ denotes projection on the leading antisymmetric module.
Its full non-linear form is of course not relevant to the differential, but the appearance of $D+\theta$ is essential for consistency, see Section \ref{sec:BV}.

This Bianchi identity can be projected on $\fk$ by contraction with 
$(G^{-1}t^\alpha)^{MN}$, which is automatically antisymmetric for $\alpha=a$ and symmetric for $\alpha=a'$.
After linearisation, one then obtains the action of the differential from torsion antifields to Bianchi identity ghost antifields, the rightmost arrow in the upper line in eq. \eqref{Cdiagram}.



\subsection{The complex}

The complex $\CC$ of extended teleparallel geometry is a modification of the gravity complex \eqref{Gdiagram}, building on the same principle.
\begin{equation}\label{Cdiagram}
    \begin{tikzcd}[row sep = 16 pt, column sep = 16 pt]
    \hbox{ghost\#}=&2&1&0&-1&-2&-3\\
        \cdots\ar[r,"d"]&V'\ar[r,"d"]&V\ar[r,"d"]&\hat\fg\ar[r,"d"]&\Theta\ar[r,"d"]&\fk \\
        &&\bar\fk\ar[r,"d",swap]\ar[ur,"\varrho" near start]&\bar\Theta\ar[r,"d",swap]\ar[ur,"\sigma" near start] 
        &\bar{\hat\fg}\ar[r,"d",swap]\ar[ur,"\varrho^\star" near start]&\bar V\ar[r,"d",swap]&\bar V'\ar[r,"d",swap]&\cdots
\end{tikzcd}
\end{equation}
Here, $\hat\fg=\fg\oplus\RR$  (generalising $\mathfrak{gl}(d)$) with $\fg$ a semi-simple Lie algebra, 
$\fk\subset\fg$ the local subalgebra, $V$ the 
generalised tangent vector space (the coordinate module), $V'$ reducibility, and $\Theta$ the torsion module
(both ``big'' and ``small''). 

The relevant parts of the differential $d$ are given in the previous subsections. The left tail, denoted by an ellipsis, is the $1$-bracket of the $L_\infty$ algebra of generalised diffeomorphisms, derived from the tensor hierarchy algebra in 
refs. \cite{Cederwall:2018aab,Cederwall:2019bai}.
It will actually contain also ancillary transformations, which are algebraic parts (\ie, without derivatives) of
the differential. They are of course essential for obtaining the correct BV action, but do not affect the arguments here, in particular, they do not participate in the condition on $\sigma$.

The central parts of the horizontal differential $d$ in the upper line can thus be written
\footnotesize
\begin{align}
&\qquad d\left(
\begin{matrix}
\ldots&V^M&\left(\begin{matrix}E^\alpha\\ e \end{matrix}\right)
	&\left(\begin{matrix}\Theta_M{}^\alpha\\ \theta_M \end{matrix}\right)
	&H_{a} 
\end{matrix}
\right)\label{eq:dGen}\\[1.5em]
&=\left(
\begin{matrix}
\ldots&\ldots&\left(\begin{matrix}t^\alpha{}_M{}^N\*_NV^M\\ \*_MV^M \end{matrix}\right)
	&\left(\begin{matrix}-\varphi^\alpha{}_{M,\beta}{}^N\*_NE^\beta\\ 
		-t_{\alpha M}{}^N\*_NE^\alpha+(\lambda,\lambda)\*_Me\end{matrix}\right)
	&\begin{matrix}(G^{-1}t_a)^{MN}\bigl(t_{\alpha\{M}{}^P\*_{|P|}\Theta_{N\}}{}^\alpha\bigr.
	\\\bigl.+2(2-{1\over(\lambda,\lambda)})\*_{\{M}\theta_{N\}}\bigr)
		\end{matrix} \label{dGen} 
\end{matrix}
\right)\;.
\nn
\end{align}
\normalsize
The action of $d$ in the lower line is (as before) given by the same ``structure constants'' between the dual modules, 
for example (the dual of eq. \eqref{eq:dEtoTheta}):
\begin{align}
\left(\begin{matrix}
\bar\Theta^M{}_\alpha\\ \bar\theta^M
\end{matrix}\right)
\longrightarrow
\left(\begin{matrix}
\bar E_\alpha\\ \bar e
\end{matrix}\right)\;:\quad
\begin{matrix}
(d\bar\Theta)_\alpha=-\varphi^\beta{}_{M,\alpha}{}^N\*_N\bar\Theta^M{}_\beta\hfill
&(d\bar\theta)_\alpha=-t_{\alpha M}{}^N\*_N\bar\theta^M\hfill\\
(d\bar\Theta)=0\hfill
&(d\bar\theta)=(\lambda,\lambda)\*_M\theta^M\hfill
\end{matrix}
\end{align}
Note that because of the convenient projection on $\fk$, $H_a$ now is a density, while $\bar H^a$ is a tensor.
The notation $\{MN\}$ denotes projection on the leading antisymmetric module, as before.
If we limit ourselves to cases when this is the only antisymmetric $\fg$-module that branches into the adjoint $\fk$-module,
$\{MN\}$ can be replaced with $[MN]$  in eq. \eqref{eq:dGen}.
In other cases, the question arises if this choice of arrow $\Theta\rightarrow H$ is correct.
We will show in Section \ref{sec:Ehlers}, dealing with the class of models where $R(-\lambda)$ is the adjoint $\fg$-module,
that it indeed is, at least in that case.

The map $\sigma$ is invertible. $\varrho$ and $\varrho^\star$ are dual maps, a subalgebra embedding:
\begin{align}
\varrho:\;&\bar H^a\mapsto\delta^a_\alpha E^\alpha\;,\nn\\
\varrho^\star:\;&\bar E_\alpha\mapsto H_a\delta^a_\alpha\;.
\end{align}
The key equation is formally the same as for teleparallel gravity,
\begin{align}
\boxed{
\;d\circ\varrho+\sigma\circ d=0\;.\;\label{eq:keypoint2}
}
\end{align}
If the lower maps are given by the same structure constants as the upper ones, 
 the right parallellogram identity $d\circ\sigma+\varrho^*\circ d=0$ is equivalent to the left one, eq. \eqref{eq:keypoint2}, as long as $\sigma$ is symmetric,
 \ie, the transformation $(\sigma\bar\Theta)_\mu=\sigma_{\mu\nu}\bar\Theta^\nu$ contains a symmetric matrix $\sigma_{\mu\nu}$. This must happen, since $\sigma$ can be diagonalised on $\fk$-modules.
 
What remains in the construction at this point is an explicit form of the dualisation $\sigma$, such that $\CC$ actually is a complex,
\ie, a solution of eq. \eqref{eq:keypoint2}.
The conditions will be purely algebraic, and the solution will not rely on the section constraint (unlike the nilpotency of the horizontal derivative). We will give explicit expressions for $\sigma$ for some classes of examples in
Sections \ref{sec:nonancillary} and \ref{sec:Ehlers}.
Since all equations are $\fk$-covariant\footnote{And in fact $\fg$-covariant, since the embedding $\fk\subset\fg$ is parametrised ``covariantly''.}, it is almost obvious that there is a solution, which can be found by diagonalising on $\fk$-modules.
It would be interesting to see if a proof of the existence of a solution (\ie, the existence of a complex) can be extended to situation with infinite-dimensional $\fg$.

A linearised BV action is built as $\langle\Psi,q\Psi\rangle$, where $\Psi\in\CC$, $q=d+\delta=d+\varrho+\sigma+\varrho^\ast$ is the sum of all arrows in  \eqref{Cdiagram}, and $\langle\cdot,\cdot\rangle$ is the natural pairing on $\CC$.

On the complex $\CC$,
homotopy transfer to the cohomology of $\delta$ can be performed. This will yield a second derivative zig-zag,
\begin{equation}
    \begin{tikzcd}[row sep = 16 pt, column sep = 16 pt]
       \cdots\ar[r]&V'\ar[r]&V\ar[r]&\fg\ominus\fk\ar[dr,out=0,in=180,looseness=4]\\
         &&&&\fg\ominus\fk\ar[r]&\bar V\ar[r]&\bar V'\ar[r]&\cdots
\end{tikzcd}
\end{equation}	
\noindent where the new arrow arises, 
according to the homological perturbation lemma \cite{Lapin}, as $d\circ\sigma^{-1}\circ d$. The procedure is the
same as in section \ref{sec:YMcomplex}.
Equivalently, one solves the algebraic equation of motion for the $\bar\Theta$ component and reinserts in the action, just like in the Yang--Mills and gravity cases.
One may also choose to transfer to the cohomology of $\sigma$, and obtain
\begin{equation}
    \begin{tikzcd}[row sep = 16 pt, column sep = 16 pt]
        \cdots\ar[r,"d"]&V'\ar[r,"d"]&V\ar[r,"d"]&\fg\ar[dr,out=0,in=180,looseness=4]&&\fk \\
        &&\bar\fk\ar[ur,"\varrho" near start]&
        &\bar\fg\ar[r,"d",swap]\ar[ur,"\varrho^\ast",near end,swap]&\bar V\ar[r,"d",swap]&\bar V'\ar[r,"d",swap]&\cdots
\end{tikzcd}
\end{equation}


\section{Examples\label{sec:examples}}

\subsection{Cases without ancillary transformations\label{sec:nonancillary}}




Let us for the moment assume that there are no subleading antisymmetric modules.
We will check the right parallellogram, $d\sigma+\varrho^\ast d=0$, and solve for $\sigma$.
The input this far gives\footnote{We sometimes switch the two pairs of incices on $\varphi$, so that it also acts ``from the left'' on the conjugate modules.} 
\begin{align}
(\varrho^\ast d\bar\Theta)_a&=-\varphi_a{}^N{}_,{}^\alpha{}_M\*_N\bar\Theta^M{}_\alpha\;,\nn\\
(\varrho^*d\bar\theta)_a&=-t_{aM}{}^N\*_N\bar\theta^M\;,
\end{align}
and
\begin{align}
(d\sigma\bar\Theta)_a=(G^{-1}t_a)^{MN}t_{\alpha M}{}^P\*_P(\sigma\bar\Theta)_N{}^\alpha\;,\nn\\
(d\sigma\bar\theta)_a=2(2-{1\over(\lambda,\lambda)})(G^{-1}t_a)^{MN}\*_M(\sigma\bar\theta)_N\;.
\end{align}

In order to make an Ansatz for $\sigma$, we first convert $(\bar\Theta,\bar\theta)$ to their conjugate modules using the involution, and write
\begin{align}
(\sigma\bar\Theta)_M{}^\alpha&=G^{\alpha\beta}G_{MN}(\sigma'\bar\Theta)^N{}_\beta\;,\nn\\
(\sigma\bar\theta)_M&=G_{MN}(\sigma'\bar\theta)^N\;.
\end{align}

Inserting into $d\sigma+\varrho^*d=0$ immediately gives the action of $\sigma'$ in the fundamental:
\begin{align}
(\sigma'\bar\theta)^M=-2(2-{1\over(\lambda,\lambda)})\bar\theta^M\;.
\end{align}
For the big torsion module, we get, after using $G^{-1}t_aG=-t_a^\transpose$, 
\begin{align}
G^{\alpha\beta}(t_at_\beta)_N{}^M\*_M(\sigma'\bar\Theta)^N{}_\beta
=-\varphi_a{}^M{}_,{}^\beta{}_N\*_M\bar\Theta^N{}_\beta\;.
\end{align}
The $\varphi$ on the right hand side acts with eigenvalues in irreducible $\fg$-modules.
The derivative can be dropped from the equation.
In order to solve this equation for $\sigma'$, it is convenient to complete it with an $a'$ part, in a way that gives a quadratic non-singular matrix on the left hand side. We do this by letting $a\rightarrow\alpha$. 
Let $x_\alpha{}^M{}_,{}^\beta{}_N=G^{\beta\gamma}(t_\alpha t_\gamma)_N{}^M$
and $y=\varphi^{-1}x\varphi$ with the obvious multiplication. 
Then, the equation is solved by the action of $\sigma'$ on $\bar\Theta$ being
\begin{align}
\sigma'=\varphi y^{-1}\;.
\end{align} 

This is the solution. If one wants to check the action on the individual $\fk$-modules, this can be done as follows.
We will sketch the procedure.
Inside the projection, 
$x_\alpha{}^M{}_,{}^\beta{}_N$ can be replaced by
$x'_\alpha{}^M{}_,{}^\beta{}_N=G^{\beta\gamma}f_{\alpha\gamma}{}^\delta t_{\delta N}{}^M$.
The operator $-f_\alpha{}^{\beta\gamma}t_{\gamma N}{}^M$ 
is known to have non-zero eigenvalues on the torsion $\fg$-modules (it is indeed a part of $\varphi$).
On a torsion module $R(\lambda+\gamma)$ they are 
\cite{Cederwall:2019qnw}
$g^\vee-g^\vee_\gamma+1$, where $g^\vee$ is the dual Coxeter number of $\fg$ and $g^\vee_\gamma$ is the dual Coxeter number of $\fg^-$, whose Dynkin diagram is obtained by deleting the node corresponding to $\lambda$.
To arrive at the eigenvalues of $\sigma'$, the only remaining piece is to calculate
\begin{align}
(\varphi^{-1})_\alpha{}^M{}_,{}^\gamma{}_P\tau_\gamma{}^\delta\varphi_\delta{}^P{}_,{}^\beta{}_N
=-P_\alpha{}^M{}_,{}^\beta{}_N
+2(\varphi^{-1})_\alpha{}^M{}_,{}^c{}_P\varphi_c{}^P{}_,{}^\beta{}_N
\;.\label{eq:phitauphi}
\end{align}
Here, one can observe that $\varphi_a{}^M{}_,{}^b{}_N$ is the ``$\varphi$-tensor'' for $\fk$ (with the same coordinate module).
The last term in eq. \eqref{eq:phitauphi} therefore gives the projection multiplied with the quotient of the
eigenvalues of $\varphi^{(\fk)}$ (for each $\fk$-module) and the eigenvalue of $\varphi$.
A complete analogous calculation of the eigenvalues of $\sigma$ will be carried out in the following subsection.

\subsection{Ehlers extended geometry\label{sec:Ehlers}}

We want to apply the construction to the case where $R(-\lambda)=\adj$ is the adjoint of $\fg$.
We view this mainly as a stepping stone towards infinite-dimensional cases.
We denote adjoint indices $M,N,\ldots$, and raise and lower freely with the Killing metric $\eta_{MN}$.
The representation matrices are $t_{MN}{}^P=-f_{MN}{}^P$.
Then,
\begin{align}
\varphi_{MN}{}^{PQ}&=2\delta_{(M}^P\delta_{N)}^Q-f_{(M}{}^{PR}f_{N)}{}^Q{}_R\;.
\end{align}
Since now, $(\lambda,\lambda)=(\vartheta,\vartheta)=2$, torsion is formed from the Weitzenb\"ock connection $(\Gamma,\gamma)$ as
\begin{align}
\Theta_{MN}&=\varphi_{MN}{}^{PQ}\Gamma_{PQ}\;,\nn\\
\theta_M&=-f_M{}^{NP}\Gamma_{NP}-2\gamma_M\;.
\end{align}
$\Theta_{MN}$ contains the subleading symmetric modules, including a scalar
$\Theta={1\over2(g^\vee+1)}\Theta_M{}^M=\Gamma_M{}^M$.

There is a notational subtlety. We define a metric $G_{MN}$, carrying some weight. As before this also leads to a metric ``$G_{\alpha\beta}$'' on the adjoint (minus the involution). However, the latter carries no weight. They are otherwise the same. 
For the calculations, we choose to simply ignore this issue. The question of weight becomes relevant when an action is formed by integrating a density; then one may simply insert the appropriate powers of $e$.

There are always Bianchi identities in $\wedge^2\adj$, which can be seen from the tensor hierarchy algebra.
This is however also a reducible module, since it contains the adjoint besides the leading antisymmetric module(s). The Bianchi identities in the leading modules are
\begin{align}
(D_P+\theta_P)(-f_{Q\{M}{}^P\Theta_{N\}}{}^Q+3\delta_{\{M}^P\theta_{N\}}^{\mathstrut})=0\;.
\label{eq:leadingBI}
\end{align}
This follows from the general case discussed earlier.
There are two Bianchi identities in the adjoint, of which one linear combination has the good $D+\theta$ form
(see Section \ref{sec:BV}):
\begin{align}
(D_N+\theta_N)(\Theta_M{}^N-{3\over2}f_M{}^{NP}\theta_P+\delta_M^N\Theta)=0\;.\label{eq:adjBI}
\end{align}
Both these Bianchi identities contain the adjoint module of $\fk$. There is only room for one $\fk$ in the complex. Which one is the right choice?
One guiding observation is that the adjoint Bianchi identity \eqref{eq:adjBI} is trivial, in the sense that the terms $\*\Gamma$ and 
$\Gamma^2$ vanish separately using the section constraint; the Maurer--Cartan equation for $\Gamma$ is not used.
Therefore, the identity becomes trivial for any (local) choice of solution to the section constraint.
The leading identity \eqref{eq:leadingBI} is needed, and the adjoint one can be added with any coefficient, giving an equivalent theory.

We now examine the right parallellogram. Denote the (local) $\fk$ indices as $a,b,\ldots$, and the complement $\fk^\perp$ 
with $a',b',\ldots$.
We have
\begin{align}
(\varrho^*d\bar\Theta)^a&=-\varphi_{PQ}{}^{aN}\*_N\bar\Theta^{PQ}\;,\nn\\
(\varrho^*d\bar\theta)^a&=f^a{}_N{}^P\*_P\bar\theta^N\;.\label{eq:rhod}
\end{align}
Note that the action of $\varphi$ in the first of these equations just returns its eigenvalues on the different modules in $\Theta$.
These are in principle known, using the methods of ref. \cite{Cederwall:2019qnw}. We illustrate their calculation in the
example with $G=E_8$, Section \ref{sec:e8example}.

Going the other way ($d\circ\sigma$), we need to extract $\fk$ from the leading antisymmetric Bianchi identity \eqref{eq:leadingBI}.
The projector on the leading antisymmetric module(s) $A$ is
\begin{align}
P^{(A)}{}_{MN}{}^{PQ}=\delta_{MN}^{PQ}+{1\over2g^\vee}f_{MN}{}^Rf^{PQ}{}_R
\end{align}
(the second term subtracts the adjoint).
Given a tensor $A_{MN}$ in $A$, $\fk$ can be extracted as $f_a{}^{bc}A_{bc}$ (or equivalently as
$f_a{}^{b'c'}A_{b'c'}$, since these sum to $0$ thanks to $f_M{}^{NP}A_{NP}=0$).
In a more ``covariant'' version, we calculate the expression implicit in eq. \eqref{eq:dGen},
\begin{align}
\Pi_a{}^{MN}&=-G^{PR}f_{aR}{}^Q P^{(A)}{}_{PQ}{}^{MN}\nn\\
&=-\bigl(G^{MP}-(1-{2h^\vee\over g^\vee})\eta^{MP}\bigr)f_{aP}{}^N\;,\label{eq:Piprojection}
\end{align}
where we in the process have used $G^{NP}f_{aN}{}^Qf_{PQ}{}^M=-2\delta_a^M(g^\vee-2h^\vee)$, which is straightforwardly derived \eg\ through splitting the indices in $\fk$ and $\fk^\perp$.
($h^\vee$ is the dual Coxeter number of $\fk$.
The notation $\delta_a^M$ means projection on $\fk$, so that $\delta_a^MV_M=V_a$.)
Applying this projection on the Bianchi identity \eqref{eq:leadingBI} then gives a way to determine $\sigma$.

$\sigma$ must act diagonally on $\fk$-modules;
acting between $\fk$-modules in $(\bar\Theta,\bar\theta)$ and their conjugate modules in $(\Theta,\theta)$.
The first thing to note is that eq. \eqref{eq:leadingBI} does not contain the scalar torsion $\Theta$,
while eq. \eqref{eq:rhod} contains the scalar $\bar\Theta$.
There are two possible ways out of this. Either one adds some non-zero constant times the $\fk$ part of the adjoint 
Bianchi identity, or one could leave the scalars out of the complex altogether. 
In the latter case, the adjoint Bianchi identity ceases to play a r\^ole in the complex---which is as well since it is trivial on any solution to the section constraint---and one is left with the leading one.
The two options are necessarily equivalent, at least after a small homotopy transfer eliminating the scalar in $\bar\Theta$.

Let us choose the second option. We forget about the scalar torsion and its dual, and consequently also the subleading antisymmetric Bianchi identity.
The traceless part of $\varphi$ is
\begin{align}
\varphi'_{MN}{}^{PQ}=\varphi_{MN}{}^{PQ}-{2(g^\vee+1)\over\dim\fg}\eta_{MN}\eta^{PQ}\;.
\end{align}
Define $\Theta'_{MN}=\varphi'_{MN}{}^{PQ}\Gamma_{PQ}$.
Using eq. \eqref{eq:Piprojection}, we obtain the conditions
\begin{align}
\bigl(G^{MN}-(1-{2h^\vee\over g^\vee})\eta^{MN}\bigr)
	f_a{}^P{}_Mf^{RQ}{}_N\*_R(\sigma\bar\Theta')_{PQ}
	&=\varphi'_a{}^M{}_{,NP}\*_M\bar\Theta'^{NP}\;,\nn\\
-3\bigl(G^{MN}-(1-{2h^\vee\over g^\vee})\eta^{MN}\bigr)f_{aM}{}^P\*_N(\sigma\bar\theta)_P
	&=-f_{aM}{}^N\*_N\bar\theta^M\;.
\end{align}

The second equation is solved by 
\begin{align}
(\sigma^{-1}\bar\theta)_M=-3\bigl(G_{MN}-(1-{2h^\vee\over g^\vee})\eta_{MN}\bigr)\bar\theta^N\;.
\end{align}
Note that the eigenvalues on $\fk$ and $\fk^\perp$ are different,
$-6(1-{h^\vee\over g^\vee})$ and $-6{h^\vee\over g^\vee}$, respectively.
In an example with $\fg= \fe_8$ and $\fk=\so(16)$ we have $g^\vee=30$, $h^\vee=14$, and the two eigenvalues are
$-{16\over5}$ and $-{14\over5}$.

In order to solve the first equation, we can drop the derivative, and extend the $a$ index to a full adjoint index.
Defining 
\begin{align}
z^{MN}{}_{PQ}=\bigl(G^{RS}-(1-{2h^\vee\over g^\vee})\eta^{RS}\bigr)
f^{(M}{}_{PR}f^{N)}{}_{QS}\;,\label{eq:zdef}
\end{align}
the solution is
\begin{align}
\sigma=\varphi'z^{-1}\;.
\end{align}
If there is a single big torsion module, $\varphi'$ can be replaced by its eigenvalue.

In order to understand the operator $z$ and how it distinguishes the different $\fk$-modules,
note that it contains two types of terms. The term contracted with $\eta$ is proportional to
the operator $-f^{(M}{}_{PR}f^{N)}{}_Q{}^R$, which on a module $R(\lambda+\gamma)$ of the type appearing in the torsion has the eigenvalue $g^\vee-g_\gamma^\vee$, where $g_\gamma^\vee=g^\vee(\fg^-)$, $\fg^-$ being the Lie algebra obtained by deleting the node of the coordinate module. $\gamma$ is a highest root with $(\lambda,\gamma)=0$.
The term contracted with $G$, containing
$-G^{RS}f^{(M}{}_{PR}f^{N)}{}_{QS}$
 works similarly. By decomposing the adjoint indices it is straightforward to 
see that it produces terms that contain the same operator, but now with respect to $\fk$. These then have the eigenvalues
$h^\vee-h_\delta^\vee$. The eigenvalues of the $-Gf\!f$ operators are
$g^\vee-g_\gamma^\vee-2(h^\vee-h_\delta^\vee)$.
Since $R(\lambda)$ splits into $\fk\oplus\fk^\perp$, both these appear as ``coordinate modules'' in the calculation.
The eigenvalues of the operator $z$ of eq. \eqref{eq:zdef} are, after a brief calculation,
\begin{align}
z_{\gamma,\delta}=2h^\vee\Bigl({g^\vee_\gamma\over g^\vee}-{h^\vee_\delta\over h^\vee}\Bigr)\;.
\end{align}

\subsubsection{An example: $E_8$\label{sec:e8example}}

Let us take an example, with $\fg=\fe_8$, $\fk=\so(16)$.
Then $g^\vee=30$, $h^\vee=14$.
The (adjoint) coordinate module is 
\begin{align}
{\bf248}=\EWeight{1000}0{00}0\;.
\end{align} 
The big torsion module is obtained from the adjoint of $\fg^-=\fe_7$, which gives 
\begin{align}
{\bf3875}=\EWeight{0000}0{01}0\;.
\end{align}
Then, $g_\gamma^\vee=g^\vee(\fe_7)=18$. The eigenvalue of the $-f\!f$ operator is $g^\vee-g^\vee_\gamma=12$ and the eigenvalue of $\varphi'$ is  
$g^\vee-g^\vee_\gamma+2=14$.
The splitting of ${\bf3875}$ into $\fk=\so(16)$ modules is obtained by considering the adjoint modules of $\fk^-$, where $\fk^-$ is obtained by deleting the node corresponding to either $\fk$ or $\fk^\perp$.
The splitting of the coordinate module is
\begin{align}
{\bf248}=\EWeight{1000}0{00}0=\DWeight{010000}00\oplus\DWeight{000000}10={\bf120}\oplus{\bf128}\;.
\end{align}
Deleting the ${\bf120}$ node gives $\fk^-=\sl(2)\oplus\so(12)$, giving rise to two highest roots $\delta_{1,2}$ with $(\theta_{\fk},\delta_{1,2})=0$.
Deleting the ${\bf128}$ node gives $\fk^-=\sl(8)$, giving one highest root $\delta_3$ with $(\Lambda_{\bf128},\delta_3)=0$.
The dual Coxeter numbers $h_{\delta_i}^\vee$ are $2$, $10$ and $8$, respectively.
Inserting the Dynkin labels of the corresponding adjoint modules into the Dynkin diagram of $\so(16)$ gives the 
splitting ${\bf3875}={\bf135}\oplus{\bf1820}\oplus{\bf1920}$, where
\begin{align}
{\bf135}&=\DWeight{200000}00\;,\nn\\
{\bf1820}&=\DWeight{000100}00\;,\\
\overline{\bf1920}&=\DWeight{100000}01\;.\nn
\end{align}
The eigenvalues of the $-G^{-1}f\!f$ operator, $g^\vee-g^\vee_\gamma-2(h^\vee-h_{\delta_i}^\vee)$, are
$-10$, $6$ and $2$, respectively, and of the $z$ operator ${64\over5}$, $-{16\over5}$ and ${4\over5}$.
The dualisation $\sigma$ thus acts by the eigenvalues
${35\over32}$, $-{35\over8}$ and ${35\over2}$ on the $\so(16)$ modules ${\bf135}$, ${\bf1820}$ and $\overline{\bf1920}$, respectively.

\section{BV actions\label{sec:BV}}

Specifying $\sigma$ gives (after ``covariantisation'') complete information of the dynamics.
In particular, the two-derivative action will, after elimination of $\bar\Theta$, contain
\begin{align}
S_0=-{1\over2}\langle \Theta(E),\sigma^{-1}\Theta(E)\rangle\;.
\end{align}
Note that the expression \eqref{eq:sigmainverse} gives the standard teleparallel gravity.

The most straightforward way of giving a full BV action of extended geometry is direct use of the complex $\CC$.
The details in the ghost sector, including ancillary ghosts, will not be spelled out here. It contains a sum of terms, 
which is infinite both in that it contains ghosts of arbitrarily high ghost number (the left tail of the complex is in general infinite), but also in that it contains brackets of arbitrarily high degree.
This part of the BV action has been constructed in full detail in ref. \cite{Cederwall:2018aab} for situations when there are no ancillary ghosts with ghost number $1$, and sketched in the presence of such ghosts in ref. 
\cite{Cederwall:2019bai}. 

Here we would like to give an account of the ``central'' part of the BV action, involving the physical fields.
The non-linear versions of Section \ref{sec:differential} of the linearised differential are used in order to construct the interacting theory, where the differential is replaced by antibracket with $S$.
There are some conventions associated with the choice normalisation of the field-antifield pairing and the BV antibracket.
We use conventions where 
$(\langle X,\Phi\rangle,\langle\bar\Phi,Y\rangle)=\langle X,Y\rangle$, 
$\bar\Phi$ being the antifield of $\Phi$ and $X,Y$ ``constants'' in appropriate modules.

The local transformations of $E$ are encoded in terms 
$\langle\bar E,E\bar H+\LL_VE\rangle$ in the BV action $S$.
There is obviously also a term $\langle\Theta(E),\bar\Theta\rangle$, where $\Theta(E)$ is the actual torsion,
the torsion part of the Weitzenb\"ock connection of $E$. (We have denoted big and small torsion collectively by $\Theta$ here.)
And there is the obvious term $\fr2\langle\bar\Theta,\sigma\bar\Theta\rangle$.
The transformation of the torsion $\Theta(E)$ under local rotations in $\fk$ is $\delta_{\bar H}\Theta(E)=D\bar H$.
Indeed, cancellation of a term $\langle\bar\Theta,D\bar H\rangle$ in $(S,S)$ arising from an antibracket
$(\langle E\bar H,\bar E\rangle,\langle\Theta(E),\bar\Theta\rangle)$
demands the covariantisation of the $\bar H\*\Theta$ term to be 
$\langle D\bar H,\Theta\rangle$.
Then the term cancels against
$(\langle D\bar H,\Theta\rangle,\fr2\langle\bar\Theta,\sigma\bar\Theta\rangle)$, using
precisely the key identity $d\varrho+\sigma d=0$.
But there is also a term 
$(\langle D\bar H,\Theta\rangle,\langle\Theta(E),\bar\Theta\rangle)
=\langle D\bar H,\Theta(E)\rangle$. This expression should vanish, expressing that the torsion Bianchi identity is an identity.

When partially integrating the covariant derivative in $\langle D\bar H,\Theta(E)\rangle$ to reveal the Bianchi identity, one needs to be careful in the presence of torsion \cite{Cederwall:2021xqi}. Remember that $\Theta(E)$ is a tensor, and carries the canonical weight $1-(\lambda,\lambda)$, while $\bar\Theta$, and thus $D\bar H$, is a density with the non-canonical weight $(\lambda,\lambda)$ (instead of the canonical $(\lambda,\lambda)-1$).
The product carries weight $1$, which is appropriate for integration, since the
naked divergence of a weight $1$ vector is covariant.
However, being covariant is not the same as containing only the covariant divergence, when the connection is torsionful.
In fact, the covariant divergence of a vector $V$ with weight $(\lambda,\lambda)$ (so that $DV$ has weight $1$) is
\begin{align}
D_MV^M=(\*_M-\theta_M)V^M\;.
\end{align}
Therefore, a total derivative is $(D_M+\theta_M)V^M$, and modulo an integral of a total derivative,
\begin{align}
\langle D\bar H,\Theta(E)\rangle=-\langle\bar H,(D+\theta)\Theta(E)\rangle\;.
\end{align}
The expression is schematic in that it does not reveal the detailed tensorial structure, but the appearance of the covariant derivative in combination with $\theta$ is precise:
the antisymmetric Bianchi identity used must be of the form
\begin{align}
(D_P+\theta_P)X_{MN}{}^P=0\;,
\end{align}
where $X$ is linear in $\Theta(E)$. This is verified in the examples in the present paper, and also in the case where 
$\fg$ is an affine Kac--Moody algebra. We expect a general proof to be possible
\cite{Cederwall:2022oyb}. 

To summarise, the part  of the BV action involving ghost number $0$ and $-1$ fields is, on a somewhat sketchy level,
\begin{align}
S&=\fr2\langle\bar\Theta,\sigma\bar\Theta\rangle+\langle\Theta(E),\bar\Theta\rangle+\langle D\bar H,\Theta\rangle\nn\\
&+\langle\bar E,E\bar H+\LL_VE\rangle+\ldots\;.
\end{align}
The master equation $(S,S)=0$ of course also requires including terms for the gauge ``algebra''
\cite{Cederwall:2018aab,Cederwall:2019bai}.
Note that the terms encoding the covariantisation of horizontal derivatives $d$, for example the second and third terms, are responsible for their appearance both in the upper line of the complex and in the lower line. 
We would like to stress again that the first term stays unmodified in the full BV theory.

\section{Conclusions and outlook}

The complex constructed in the present paper provides a clear way to define (teleparallel) dynamics of
extended geometry with structure algebra $\fg$. The essential information is carried by the dualisation $\sigma$, and it suffices to find an expression for $\sigma$ in the linear model, as it remains undeformed in the full theory.
The only property one needs to verify for the non-linear model is that there is a torsion Bianchi identity of the correct form,
where the covariant derivative occurs with $D_M+\theta_M$. We believe that a general proof of this property may be formulated using the methods of ref. \cite{Cederwall:2022oyb}.
As an attractive byproduct, it becomes clear how a BV action is formulated, using input from the tensor
hierarchy algebra $S(\fg^+)$.

It will be interesting to see if we can apply the systematics presented here to extended geometry based on infinite-dimensional structure groups. 
Affine extended geometry has been constructed \cite{Bossard:2017aae,Bossard:2018utw,Bossard:2021jix}, and partial result exist
on very extended structure groups \cite{Bossard:2021ebg}. We believe that the present formalism, with its close connection to tensor hierarchy algebras, will provide the most efficient and clear-cut method to construct such models.
The relevant tensor hierarchy algebras are under reasonable control \cite{Cederwall:2021ymp,Bossard:2021ebg}.
One immediate difficulty is that the locally realised involutory subalgebras 
\cite{Kleinschmidt:2006dy,Kleinschmidt:2018hdr,Kleinschmidt:2021agj}
are not Kac--Moody algebras, and little is known about branching of highest weight modules of $\fg$ into $\fk$-modules. Progress on this issue is desirable.

\subsection*{Acknowledgments} The authors would like to thank Michael Reiterer, Olaf Hohm, Ingmar Saberi and Guillaume Bossard
for stimulating discussions and input.
Part of this work was done during the workshop on Higher Structures, Gravity and Fields at the Mainz Institute for Theoretical Physics
of the DFG Cluster of Excellence PRISMA${}^+$ (Project ID 39083149).
We thank the institute for its hospitality.

\bibliographystyle{utphysmod2}


\providecommand{\href}[2]{#2}\begingroup\raggedright\endgroup

\end{document}